\documentclass[aps,prl,twocolumn,superscriptaddress,longbibliography]{revtex4-2}

\usepackage{mathrsfs}
\usepackage{tabularx}
\usepackage{slashed}
\usepackage{amsmath}
\usepackage{amsfonts}
\usepackage{graphicx,xcolor}
\usepackage{times}
\usepackage[caption=false]{subfig}
\usepackage[colorlinks,linkcolor=red,citecolor=blue]{hyperref}
\usepackage{amsthm}
\usepackage[normalem]{ulem}

\begin{document}

\title{Stable boundary modes for fragile topology from spontaneous PT-symmetry breaking}

\author{Kang Yang}
\affiliation{Dahlem Center for Complex Quantum Systems, Fachbereich Physik, and Halle-Berlin-Regensburg Cluster of Excellence CCE, Freie Universit\"at Berlin, 14195 Berlin, Germany}
\affiliation{Department of Physics, School of Science, Westlake University, Hangzhou 310030, China}
\author{Fei Song}
\affiliation{Kavli Institute for Theoretical Sciences, Chinese Academy of Sciences, 100190 Beijing, China}
\author{Piet W. Brouwer}
\affiliation{Dahlem Center for Complex Quantum Systems, Fachbereich Physik, and Halle-Berlin-Regensburg Cluster of Excellence CCE, Freie Universit\"at Berlin, 14195 Berlin, Germany}

\date{January 2026}

\begin{abstract}
Two-dimensional topological insulators protected by nonlocal symmetries or with fragile topology usually do not admit robust in-gap edge modes due to the incompatibility between the symmetry and the boundary. Here, we show that in a parity-time (PT) symmetric system robust in-gap topological edge modes can be stably induced by non-Hermitian couplings that spontaneously break the PT symmetry of the eigenstates. The topological edge modes traverse the imaginary spectral gap between a pair of fragile topological bands, which is opened by the presence of the non-Hermitian perturbation. We demonstrate that the net number of resulting in-gap modes is protected by an operator version of anomaly cancellation that extends beyond the Hermitian limit. The results imply that loss and gain can in principle drive fragile topological phenomena to stable topological phenomena.
\end{abstract}

\maketitle

{\em Introduction.---}Anomalous in-gap boundary modes --- modes that cannot be created or removed by local perturbations --- are highly efficient tools for the diagnosis of topological phases \cite{RevModPhys.82.3045,RevModPhys.83.1057,lu2014topological,annurev-conmatphys-031214-014501,RevModPhys.89.025005}. These modes appear on the boundaries of topological systems and on the interfaces between systems that hold different topological indices. 
Robust in-gap boundary modes do not appear for all topological phases: They appear for stable topological phases only and
usually require invariance of the boundary under the symmetry protecting the system. This condition implies that many topological states relying on nonlocal symmetries and fragile topological phases may not be identified via boundary signatures \cite{PhysRevB.82.241102,PhysRevLett.106.106802,PhysRevB.83.245132,hsieh2012topological,science.aah6442,PhysRevLett.119.246402,PhysRevX.9.031003}.  

Recently, topology protected by parity-time (PT) symmetry has received considerable attention \cite{PhysRevLett.116.156402,PhysRevLett.118.056401,PhysRevB.95.235425}. Invariance with respect to the combined action of spatial inversion (P) and time-reversal (T) can impose nontrivial topology onto a pair of bands, which is indexed by the {\em Euler number} \cite{PhysRevX.9.021013,bouhon2020non,PhysRevLett.125.126403}. A nonzero Euler number of the two central bands in twisted bilayer graphene is known to form an obstruction to a Wannier basis involving these bands only \cite{PhysRevB.98.085435,PhysRevX.9.021013,PhysRevB.99.195455,PhysRevLett.123.036401}. Euler bands have been reported to lead to rich correlated physics \cite{PhysRevX.8.031089,PhysRevLett.122.106405,PhysRevLett.129.047601,PhysRevB.106.245129,ledwith2024nonlocal,kolavr2025robustness} in moir\'e materials and the Euler number of a pair of bands is closely related to the non-Abelian phenomena when band touching points are braided around each other \cite{wu2019non,bouhon2020non,PhysRevLett.125.053601}. Nevertheless, the Euler topology is fragile in the sense that hybridizing Euler bands with a trivial band removes the topology \cite{PhysRevLett.121.126402,PhysRevLett.125.126403,PhysRevB.108.155137}. As a result of the fragility of the Euler topology, there are no directly measurable consequences, such as robust boundary states \cite{PhysRevB.100.205126,science.aaz7650,PhysRevLett.133.093404,PhysRevLett.133.186601,WU20243657,PhysRevB.111.L081103}.

\begin{figure}
    \centering
    \includegraphics[width=\linewidth]{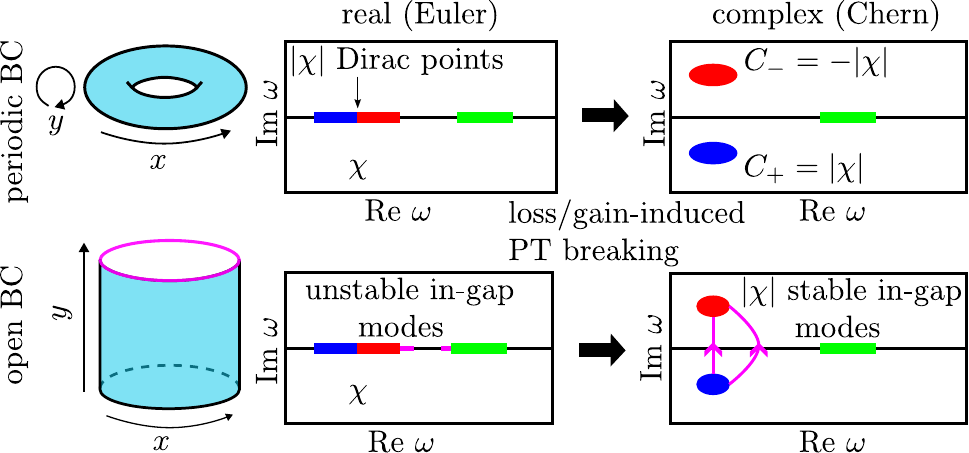}
    \caption{Top: In the spontaneous PT-breaking transition, a pair of real bands transitions into a pair of complex conjugate bands without hybridizing with any other bands. The topological invariant of the real band pair with PT symmetry is the Euler number $\chi$, featuring $|\chi|$ unremovable Dirac points between the pair \cite{PhysRevX.9.021013}. The complex bands arising from the spontaneous breaking of PT symmetry have Chern numbers $C_{\pm} = \pm \chi$ \cite{yang2025spontaneous}. Bottom: With open boundary conditions, the complex bands are linked via in-gap boundary modes (magenta). In a cylinder geometry (periodic boundary conditions in $x$ direction, open boundary conditions in $y$ direction), the boundary modes exhibit spectral flow, such that the total number of boundary modes, weighed with their spectral flow direction, equals $|C_{\pm}| = |\chi|$.}
    \label{fig_dem}
\end{figure}

The situation is fundamentally different in non-Hermitian PT-symmetry systems. In non-Hermitian PT-symmetric systems, PT symmetry may be spontaneously broken on the level of individual eigenvalues and eigenvectors  \cite{Bender_2007,PhysRevLett.103.093902,ruter2010observation,PhysRevLett.106.093902,PhysRevX.4.031042}. In this spontaneous PT-breaking transition, a pair of real bands morph into a pair of bands with complex conjugate eigenvalues and eigenstates, without hybridizing with any of the other bands, as shown schematically in Fig.\ \ref{fig_dem}. If the PT-breaking transition is complete for the band pair--- which means that the resulting complex bands are separated by an imaginary spectral gap ---, the individual complex band carry stable Chern topology inherited from the fragile topology of the  real Euler band pair, $C_{\pm} = \pm \chi$, 
a phenomenon referred to as the ``Chern-Euler duality principle'' \cite{yang2023homotopy, yang2025spontaneous}. Via this real-complex  transition, non-Hermiticity provides a pathway to reach stable PT-symmetric topological phases, without closing the gap to any of the other bands in the spectrum. The spontaneous breaking of PT symmetry is particularly relevant for PT-symmetric systems implemented in photonic and acoustic platforms \cite{jiang2021experimental,PhysRevX.13.021024,xue2023stiefel,PhysRevB.108.085116,science.adf9621,JIANG2024,PhysRevLett.132.197202}, where effective non-Hermiticity naturally arises from inevitable gain and loss \cite{feng2017non,Longhi_2017,el2018non,ozdemir2019parity,RevModPhys.91.015006}.

In this work, we show that non-Hermiticity can be leveraged to endow a pair of Euler bands with a robust unique boundary signature: The Chern-Euler duality transition
produces in-gap boundary modes linking the two complex bands. If an imaginary spectral gap is opened between the pair of fragile topological bands for periodic boundary condition, it is also opened between the band pair in the presence of boundaries. Hereby each in-gap boundary mode can be assigned a spectral flow direction, such that the number of boundary modes, counted according to the direction of their spectral flow, equals the Chern number $C_{\pm}$ and, hence, is related to $\chi$ via the Chern-Euler duality. The presence of a net spectral flow means that, unlike eventual boundary modes of the real Euler bands, which are non-topological and can always be removed by a suitably chosen local perturbation \cite{PhysRevLett.125.126403}, the in-gap boundary modes in the PT-broken case are topologically robust, in the sense that they cannot be absorbed into the bulk by local perturbations on the boundaries.


\begin{figure*}
    \centering
    \includegraphics[width=\linewidth]{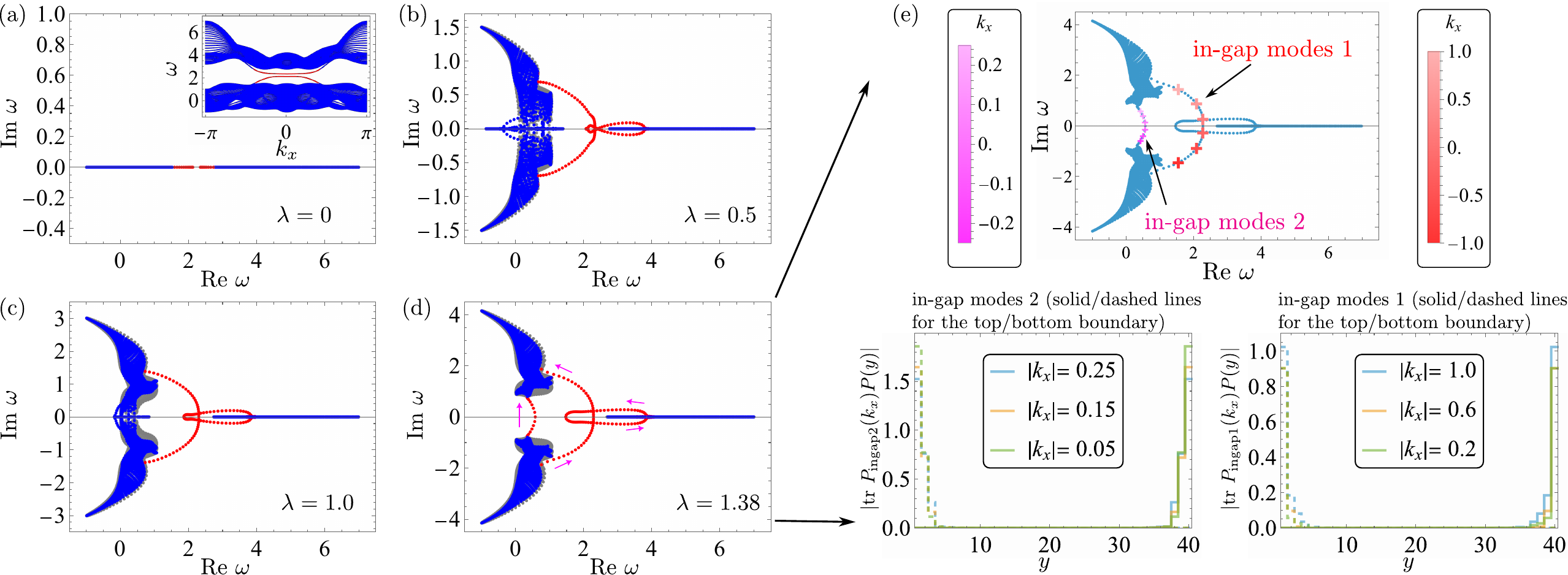}
    \caption{The in-gap edge modes (in red color) during the spontaneous symmetry-breaking transition, computed at $L=40$. (a) In the Euler-band model, the system exhibits trivial edge modes at each boundary. (b)-(c) Adding anti-Hermitian terms lifts the in-gap modes to the complex plane. (d) After the bulk modes are lifted away from the axis, in-gap modes connecting the two bulk spectra appear. The spectral flow of the top-boundary ($y=L$) in-gap modes indicated in magenta arrow. The bottom-boundary in-gap modes ($y=0$) are degenerate with the right-boundary modes and carry opposite spectral flow. (e) Top: the momentum-dependence of the in-gap modes at the top boundary, which gives the spectral flow in (d). Bottom: the biorthogonal localization $|\langle y
|\psi_\textrm{ingap}(k_x)\rangle\langle\bar\psi_\textrm{ingap}(k_x)|y\rangle|=|\textrm{tr }P_\textrm{ingap}(k_x)P(y)|$ of the top-boundary in-gap modes (solid) as well as the localization of their PT partners---the bottom boundary modes (dashed).}
    \label{fig_EdTran}
\end{figure*}

{\em Case study: Boundary modes for the Chern-Euler duality transition in a three-band model.---}
A minimal model exhibiting the Chern-Euler duality transition is a spinless model defined on the square lattice, with three orbitals per lattice site. The basis states are denoted $|\mathbf r,a\rangle$, where the orbital index $a = 1,2,3$. 
In Fourier space, PT reversal amounts to $|\mathbf k,a\rangle\to |\mathbf k,a\rangle^\ast$, so that the Bloch Hamiltonian $H(\mathbf k)$ of a PT-symmetric system is a real matrix. We allow for gain and loss processes, so that the $3 \times 3$ matrix $H(\mathbf k)$ does not need to be symmetric. 

To describe spontaneous PT-symmetry breaking, we consider a one-parameter family of PT-symmetric matrices 
\begin{equation}
  H_{\lambda}(\mathbf k) = H_0(\mathbf k) + \lambda H_1(\mathbf k),
  \label{eq:Hlambda}
\end{equation}
where we add the real antisymmetric matrix $H_1(\mathbf k)$ to the real symmetric matrix $H_0(\mathbf k)$ to induce symmetry breaking in the eigenvectors. Specifically, we choose a model with nearest-neighbor and next-nearest-neighbor hopping defined by
\begin{align}
  H_0(\mathbf k)=&\,
    \begin{pmatrix}
 g_1(k_x,k_y) & f_1(k_x,k_y) & f_2(k_x)\\
 f_1(k_x,k_y) & g_1(k_y,k_x) & f_2(k_y)\\
 f_2(k_x) & f_2(k_y) & g_2(k_x,k_y)
    \end{pmatrix}+\nonumber\\
    &\delta|n(\mathbf k)\rangle\langle n(\mathbf k)|, \\
  H_1(\mathbf k) =&\,
  \begin{pmatrix}
    0 & h(k_x,k_y) & \gamma+\sin k_y \\
    -h(k_x,k_y) & 0 & - \sin k_x \\
    -\gamma- \sin k_y & \sin k_x & 0
  \end{pmatrix}, 
  \label{eq:H1}
\end{align}
where $g_1(k_x,k_y)=\cos k_y + 2\sin^2 k_x$, $g_2(\mathbf k)=2 \sin^2 k_x + 2 \sin^2 k_y -0.6(\cos k_x+\cos k_y)$, $f_1(k_x,k_y)=2 \sin k_x \sin k_y$, $f_2(k)=-0.6 \sin k+\sin 2k$, and $h(k_x,k_y) = 1 - \cos k_x - \cos k_y$. The (bulk) spectrum of $H_0(\mathbf k)$ consists of two real lower-energy bands and one higher-energy band separated by a spectral gap; see Fig.\ \ref{fig_EdTran}a. There is no spectral gap between the two lower-energy bands, which together carry an Euler number $\chi = 2$. (The model $H_0(\mathbf k)$ is a minimal model with nontrivial Euler topology, since the Euler number must always be even for a three-band model \cite{PhysRevB.102.115135}.) In contrast, the spectrum of $H_1(\mathbf k)$ is purely imaginary, with one flat band at eigenvalue $\omega = 0$ and two complex conjugate bands separated by an imaginary gap. The nonreciprocal coupling $\gamma=0.2$ in $H_1$ can switch on non-Hermitian skin effects \footnote{The model at $\gamma=0$ is also symmetric with respect to P and T separately; The presence of space inversion symmetry makes non-Hermitian skin effects absent, more details in \cite{suppm}.}. 
Starting from $\lambda=0$, the two-lower energy bands of $H_0(\mathbf{k})$ transition into two complex conjugate bands upon gradually tuning $\lambda$ to non-zero values, while preserving a spectral gap to the third band. 
Since the third band does not participate in the real-complex transition, we refer to it as the ``remote band''. The size of the spectral gap to the remote band can be further tuned by the term $\delta|n(\mathbf k)\rangle\langle n(\mathbf k)|$ in $H_0$, where $n(\mathbf k)=(-\sin k_x,-\sin k_y,h(k_x,k_y))$ and we fix $\delta=0.2$.

To investigate boundary states associated with the spontaneous real-complex transition, we consider the model
of Eqs.\ (\ref{eq:Hlambda})--(\ref{eq:H1}) on a cylinder geometry with finite size $1 \le y \le L$ and periodic boundary conditions with period $L_x$ along $x$. Spectra for $\lambda = 0$, $0.5$, $1.0$, and $1.38$ are shown in Fig.\ \ref{fig_EdTran}. The bulk states are colored blue. The same system on the torus geometry is plotted in gray, demonstrating the influence of skin effects. The in-gap states are marked in red, localized either on the bottom boundary $y=1$ or the top boundary $y=L$. This can be explicitly visualized through biorthogonal localization \cite{PhysRevLett.121.026808} in the presence of skin effects, which is essentially the overlap between the state projector $P_{\textrm{ingap}}=|\psi_{\textrm{ingap}}\rangle\langle\bar\psi_{\textrm{ingap}}| $ ($\langle\bar\psi|,|\psi\rangle$ as the biorthogonal left and right eigenstates) \cite{kato2013perturbation} and the $y$-coordinate projector $P(y)=|y\rangle\langle y|$, as shown in Fig.\ \ref{fig_EdTran}e. The in-gap states at opposing boundaries are degenerate and their degeneracy can be lifted by including an Hermitian inversion breaking term \cite{suppm}. The two complex bands are fully separated at $\lambda \approx 1.3$ for both cylinder and torus geometry (In fact, the imaginary spectral gaps for both boundary conditions are opened simultaneously despite skin effects, see Appendix A). We can prove that the resulting complex bands carry Chern numbers $C_\pm=\pm\chi$ respectively, by generalizing the approach of Ref.~\onlinecite{yang2025spontaneous} to real-space band projectors (details in Appendix B).

To visualize the spectral flow, we consider the dependence of the boundary mode energies on $k_x$, which takes the discrete values $(2 \pi n + \varphi)/L_x$, with $n$ integer, if an Aharonov-Bohm flux $\varphi$ is applied along $y$. Upon adiabatically increasing $\varphi$ by $2 \pi$, the discrete eigenvalues of the modes at the top boundary at $y=L$ are shifted by one spacing in the direction on of the arrows in Figs.\ \ref{fig_EdTran}d, e. (The eigenvalues of the PT-inverted modes at the bottom boundary at $y=1$ are shifted in the opposite direction.) The boundary modes localized near $y=L$ have spectral flow pointing away from the bulk band with $\mbox{Im}\, \omega < 0$ and Chern number $C_+ = 2$ towards the bulk band with $\mbox{Im}\, \omega > 0$ and Chern number $C_- = -2$. The total flow carried by these modes between the two complex bands is equal to $|C_\pm|=|\chi|=2$. 

We show that the spectral flow is robust against boundary perturbations. In contrast, the two other features in Fig.\ \ref{fig_EdTran} are accidental and not rooted in the dual Chern-Euler topology: (i) the in-gap boundary modes already existing before the real-complex transition, {\em e.g.}, for $\lambda = 0$, see Fig.\ \ref{fig_EdTran}a, (ii) after the real-complex transition, the connection between the complex bands and the remote real band via in-gap boundary modes..

\begin{figure}
    \centering
    \includegraphics[width=\linewidth]{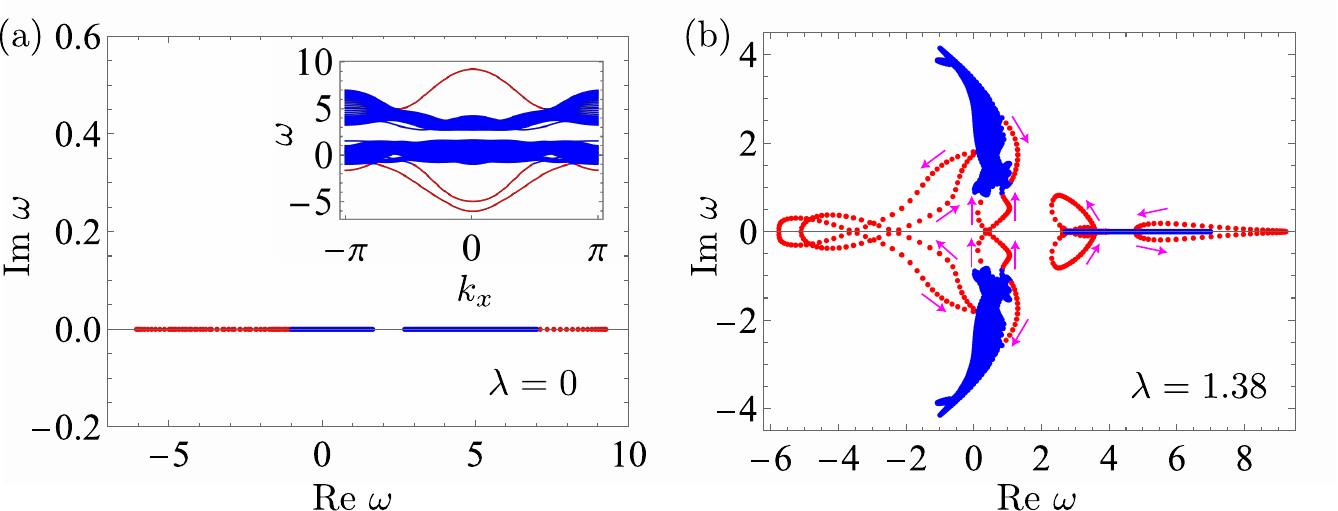}
    \caption{Spectra of the systems under boundary perturbations. (a) The in-gap modes for the Euler bands can be pushed into the bulk by the boundary perturbation. (b) After the symmetry breaking, the in-gap modes connecting the complex bands persist and the net flow remains unchanged.}
    \label{fig_bdpt}
\end{figure}

To demonstrate this, we include a PT-symmetric boundary perturbation $H_{\mathrm {bd}}$ acting locally at the top and bottom boundaries,
\begin{align}
  H_{\mathrm {bd}}(k_x)=&-\zeta_1(k_x)|k_x,L,1\rangle\langle k_x,L,1|
  -\zeta_2(k_x)\left(|k_x,L,2\rangle\right.\nonumber\\&\left.\langle k_x,L-1,2|+ |k_x,L-1,2\rangle\langle k_x,L,2|\right)+\nonumber\\ &\textrm{PT partners},
  \label{eq:Hbd}
\end{align}
where $|k_x,y,a\rangle$ is the $x$-direction Fourier transformation of the states $|x,y,a\rangle$ and $\zeta_1(k_x)= 3+3\cos k_x\,,\zeta_2(k_x)=2.8+4.2\cos k_x$. This boundary perturbation pushes the two in-gap modes in the Euler model $H_0(\mathbf k)$ back into the bulk spectrum as shown in Fig.~\ref{fig_bdpt}a, demonstrating their fragility. The boundary modes attached to the complex bulk bands for finite $\lambda$, when $H_{\lambda}(\mathbf k)$ is non-Hermitian, are not removed by the perturbation $H_{\mathrm {bd}}$. For $\lambda = 1.38$, when the two complex bands are well separated by an imaginary spectral gap, the in-gap modes emerging from the complex bands are now completely detached from the remote band. The perturbed model has more in-gap boundary modes than the unperturbed model (compare Figs.\ \ref{fig_EdTran}d and \ref{fig_bdpt}b). However, the additional modes do not carry net spectral flow. This indicates a topological protection for the boundary spectral flow between the complex bands. 


{\em Bulk spectral flow in non-Hermitian regime.---} The topological robustness of the in-gap modes is connected to the Chern numbers $|C_\pm|=|\chi|$ carried by the complex bands. In quantum Hall physics, in-gap edge modes carrying spectral flow must appear due to an equivalent bulk spectral flow given by the Chern number, known as the anomaly cancellation mechanism \cite{STONE199138,RevModPhys.89.025005}. We now demonstrate its applicability in the non-Hermitian regime using an operator formalism \cite{KITAEVHoneycomb}, and thus prove that the net number of the in-gap modes is related to the Chern number of the bulk bands.

We first address the bulk spectral flow for periodic boundary conditions along $x$ and $y$, where only bulk states exist and the flow can be clearly seen. As the flux insertion shifts the momentum $k_x$ and creates flow in the $y$-direction, it is convenient to introduce the mixed form $3L\times 3L$ Hamiltonian $H_{t,s}(k_x)$ ($s,t$ are real-space $y$-coordinates and we leave orbital indices implicit) and the operator $P_{t,s}(k_x)$ that project to one of the Chern bands. Formally, $P_{t,s}(k_x)$ is defined as \cite{kato2013perturbation}
\begin{align}
  P_{t,s}(k_x)& = \frac{1}{2 \pi i} \oint_{\mathcal C_\alpha} dz[z-H(k_x)]^{-1}_{\quad t,s}
  \label{eq:Pdef}
\end{align}
where $\mathcal C_\alpha$ is a contour enclosing the spectrum of the Chern band on the complex plane. The projection $P$ is an oblique projection, $P\ne P^\dagger$. Away from exceptional points $P$ can alternatively be expressed through the biorthogonal eigenstates of the complex band \cite{kato2013perturbation}. For periodic boundary conditions, it is related to the $3\times 3$ Bloch-state projector via Fourier transformation: $P_{t,s}(k_x)=\int dk_y\tilde P(\mathbf k)\exp[ik_y(t-s)]/(2\pi)$, where $\tilde P(\mathbf k)= \oint_{\mathcal C_\alpha} dz[z-H(\mathbf k)]^{-1}/(2\pi i)=|\psi(\mathbf k)\rangle\langle\bar\psi(\mathbf k)|$. With this relation, 
the band projector can be shown to be translationally invariant $P_{s,t}(k_x) = P_{s-t}(k_x)$ and decay quickly (decay length denoted as $R$) as a function of $|s-t|$ due to the spectral gaps to other bands \cite{PhysRev.115.809,PhysRevLett.82.2127,RevModPhys.71.1085,PhysRevLett.86.5341,PhysRevB.74.235111,demko1984decay}. 

The number of states associated with a band $\alpha$ is $N_\alpha(k_x)=\textrm{tr } P(k_x)$, the rank of the projector. We partition this number into its components inside and outside a strip of the bulk. 
To achieve this, we introduce a diagonal operator $Q$ that takes value $1$ for $y_1\le y\le y_2$ and value $0$ else. The operator satisfies $Q^2=Q$ and projects any state onto its components on $y_1\le y\le y_2$. The width of the strip $|y_2-y_1|$ is taken to be larger than the decay length $R$ of $P_{s,t}(k_x)$. The total number of states inside a band is divided into $\textrm{tr }P=\textrm{tr }PQ+\textrm{tr }P(I-Q)=n_{\rm in}+n_{\rm out}$. As $P$ is an oblique projection, $n_{\rm in}$ and $n_{\rm out}$ are not guaranteed to be real. 

We look at how many states pass through the strip when an Aharonov-Bohm flux $\varphi = 2 \pi$ is inserted. This is given by
\begin{equation}
    \int dk_x\delta n_{\rm in}(k_x)=\int dk_x\ \textrm{tr } Q\dot{P}(k_x),
\end{equation}
where we adopt the notation $\dot{P}(k_x)\equiv dP(k_x)/dk_x$. By differentiating $P^2=P$, we have $\dot{P}=-[[P,\dot{P}],P]$.
The flow of states is then expressed as
\begin{equation}
  -\delta n_{\rm in}
  =\textrm{tr}\left([P,\dot P][P,Q]\right)\label{eq_cc}
\end{equation}
Expanding the commutator $[P,Q]_{t,s}=P_{t,s}[\theta(s-y_1)\theta(y_2-s)-\theta(t-y_1)\theta(y_2-t)]$, where $\theta$ is the Heaviside step function, we can see that $[P,Q]$ is nonzero only when one of $s,t$ is inside the strip and the other outside the strip. Together with the decay property of $P_{t,s}$ for large $|t-s|$, we reduce Eq.~\eqref{eq_cc} to contributions near the strip boundaries $\delta n_{\rm in}\approx J_{y_2}-J_{y_1}$, where
\begin{align}
    J_{y_j}=&\sum_{s,t\sim y_j}\textrm{tr }[P,\dot P]_{s,t}P_{t,s}[\theta(s-y_j)-\theta(t-y_j)]\nonumber\\
    =&\sum_{|r|<R}\textrm{tr }[P,\dot P]_{r} P_{-r}
    \sum_{t\sim y_j}[\theta(t+r-y_j)-\theta(t-y_j)]\nonumber\\
    \approx&\sum_{|r|<\infty}\textrm{tr }[P,\dot P]_{r}P_{-r} r
    =\int\frac{dk_y}{2\pi i}\textrm{tr }\tilde P[\partial_{k_x}\tilde P,\partial_{k_y}\tilde P].
\end{align}
In the first line, the summation over $s,t$ is restricted to a neighborhood of $y_j$ of width $R$. In the second line, we relabeled $s$ as $r+t$ and made use of the translational invariance, writing $P_{t+r,t}$ as $P_r$. As $|P_{r}P_{-r}r|$ decays rapidly (which follows from the analyticity of the Fourier transform $\tilde P(\mathbf k)$), the summation of $r$ in the third line can be extended to infinity. The integral of $J_{y_j}$ over $k_x$ is then equal to the Chern number $C_\alpha$.

The above equations indicate a persistent flow of states in the $y$ direction inside the bulk of the system upon inserting a flux quantum parallel to $y$. We now establish the existence of the same spectral flow for a system with open boundary conditions for the $y$ coordinate. In this case, the flow is perpendicular to the boundary, where it must be diverted to another band. Because the complex bands are separated from other bands by bulk spectral gaps, the flow to other bands can only take place via boundary modes that reside in the spectral gaps of the bulk modes. This implies the existence of $|C_\pm|=|\chi|$ boundary modes that connect the complex bands. 

We extend the spectral flow calculation to the case of open boundary conditions along $y$. The projection $P_{t,s}(k_x)$ is still defined with Eq.\ (\ref{eq:Pdef}). In the presence of skin effects, we cannot immediately relate it to the conventional Bloch state projector. But, as observed in previous literature \cite{PhysRevLett.121.086803,PhysRevLett.116.133903,PhysRevLett.121.136802,PhysRevLett.121.026808,PhysRevLett.123.246801,PhysRevLett.126.077201,wang2022non}, the two key properties used in the derivation of the spectral flow can still hold: (a) The band projector $P_{s,t}(k_x)$ is asymptotically translationally invariant, $P_{s,t}(k_x) \approx P_{s+t',t+t'}(k_x)$, for coordinates $s$ and $t$ sufficiently far from the boundary, and (b) $P_{s,t}(k_x)$ decays quickly as a function of $|s-t|$ due to the spectral gaps. The two properties can be derived with the non-Bloch band theory \cite{boettcher2005spectral,PhysRevLett.123.066404,PhysRevLett.125.226402,PhysRevX.14.021011} (details in Appendix A). 
The translational invariance and rapidly decaying property lead to a continuous Fourier transformation $\tilde P(\mathbf k)\equiv\sum_r P_{r,r_0}(k_x)\exp[-ik_y(r-r_0)]$ (for any reference $r_0$ well inside the bulk $1 \ll r_0 \ll L$) under open boundary conditions. In non-Bloch band theory \cite{PhysRevLett.123.066404,PhysRevLett.125.226402,PhysRevX.14.021011}, the momentum $k_y$ sometimes needs to be shifted by an imaginary part $i\mu$ to obtain a convergent $\tilde P(\mathbf k)$. However, for the PT-symmetric systems we considered, the rapid decaying $P_{r,r_0}(k_x)$ makes real $k_y$ sufficient (see Appendix A). Moreover, the continuity properties of the projectors during the real-complex transition guarantee that $\tilde P(\mathbf k)$ defined from the Fourier transformation of the real-space projector $P_{t,s}(k_x)$ also carries $C_\pm=\pm|\chi|$ (see Appendix B). The remainder of the construction of the spectral flow  is then identical to the calculation for periodic boundary conditions along $y$.

{\em Discussion.---}
The above demonstration can be directly generalized to pairs of fragile topological bands of arbitrary Euler index in any $N$-band system. Thus, we show a universal mechanism to generate robust in-gap modes that quantify fragile parity-time and $C_2T$ symmetric topological systems. Under the Chern-Euler duality,  in-gap modes carrying spectral flow must appear after the symmetry breaking transitions induced by non-Hermitian couplings. They transfer $|\chi|$ states between the attenuated bands and the amplified bands, which may be detected by boundary pumping methods \cite{zilberberg2018photonic}. 

As the number of in-gap modes is given by the Euler number, this indicates that loss and gain can help to diagnose fragile topological states on photonic and acoustic platforms. In practice, the mechanism will be most conveniently implemented in fragile bands with narrow bandwidths, where the imaginary spectral gap can be opened already for small non-Hermitian couplings. Examples include band structures produced by stacking Dirac points via moir\'e patterns, which may be engineered in optical lattices \cite{esparza2024exceptional} and photonic devices \cite{wang2020localization,PhysRevResearch.4.L032031}. 

The generation of in-gap edge modes is robust against skin effects. Interplays with other boundary conditions and higher-order skin effects \cite{PhysRevLett.123.016805,PhysRevB.102.205118} will be interesting for future studies.

{\em Acknowledgements.---}
The authors acknowledge discussions with Ken Shiozaki, Wojciech Jankowski, Tom\' a\u{s} Bzdu\u{s}ek, Sachin Vaidya, and Zhong Wang. We also acknowledge Zhi Li, Peng Xue, and Emil Bergholtz for collaboration on preceding work. P.W.B. was supported by the Deutsche Forschungsgemeinschaft (DFG, German Research Foundation) - Project Number 277101999 - CRC-TR 183 (projects A02 and A03). F.S. acknowledges supports from NSFC under Grant No.~12404189 and from the Postdoctoral Fellowship Program of CPSF under Grant No. GZB20240732 .

\bibliography{CEBD}

\section{Appendix A: Properties of band projectors}

We consider a band $\alpha$ (or a set of bands) with well-developed spectral gaps $|\Delta_{\alpha,\beta}|=\textrm{inf}_{i\in \alpha,j\in\beta}|\omega_i-\omega_j|>0$ to other bands $\beta\ne\alpha$, that is, the gaps exist for both periodic boundary conditions (PBCs) and oppen boundary conditions (OBCs). As the open-boundary spectra lie within the periodic-boundary spectra \cite{boettcher2005spectral,PhysRevX.14.021011}, this assumption can be simplified to the existence of spectral gaps for PBCs only. We make this assumption for two reasons. First, this is a practical assumption of unambiguous spectral gaps in reality, as an exponentially small coupling $\sim\exp(-L)$ between the boundaries of the system can switch the bulk spectrum to PBC results \cite{boettcher2005spectral}. The realistic bulk spectrum could be in between the PBC spectrum and the OBC spectrum. Second, as we will show in the end of this appendix, the imaginary spectral gap between the band pair opens simultaneously for PBC and OBC cases. So this assumption is satisfied in the PT-breaking transition. We show that such band projector $P_{t,s}(k_x)$ is translationally invariant and decays quickly in real space. Here in the whole appendix, the symbol $P_{t,s}(k_x)$ can refer to any band(s), not limited to the  Chern bands in the main text.

In the absence of skin effects, the property of band projection operators can be obtained from the Bloch theory:
\begin{equation}
    P_{t,s}(k_x)\approx\int \frac{dk_y}{2\pi} u_{k_x}(k_y)e^{ik_y(t-s)},
\end{equation}
where $\approx$ means up to corrections near the boundary, and $u_{k_x}(k_y)$ is actually the Bloch-state projection operator for this case. The translational invariance is evident. When $u_{k_x}(k_y)$ is a $C^n$-function in $k_y$, we can perform integrals by parts
\begin{align}
    |P_{t,s}(k_x)|\approx& \left\vert\left(\frac{i}{t-s}\right)^n\int\frac{dk_y}{2\pi}\frac{d^nu_{k_x}(k_y)}{d^nk_y}e^{ik_y(t-s)}\right\vert\nonumber\\
    \le &\frac{1}{|t-s|^n} \int\frac{dk_y}{2\pi}\left\vert\frac{d^nu_{k_x}(k_y)}{d^nk_y}\right\vert.
\end{align}
For analytic $u_{k_x}(k_y)$, $P_{t,s}$ decays faster than any power laws. 

A typical class of skin effects comes from a non-unitary transformation on the real-space basis $S:|s\rangle\to e^{\mu s}|s\rangle$. This transformation acts on the left eigenvectors inversely. Under the non-unitary transformation, the band projection operator takes the form:
\begin{align}
    P_{t,s}(k_x)\approx&\int \frac{dk_y}{2\pi} u_{k_x}(k_y)e^{(ik_y-\mu)(t-s)}\nonumber\\
    = &\int \frac{dk_y}{2\pi} u_{k_x}(k_y-i\mu)e^{ik_y(t-s)},
\end{align}
where we shift the integral by $i\mu$ in the second line. The previous trick still applies if $u_{k_x}(z)$ is analytic on the strip of width $|\mu|$ around the real axis. The singularities of $u_{k_x}(z)$ on the complex plane are related to the spectral gaps to other bands \cite{PhysRevLett.82.2127,RevModPhys.71.1085,PhysRevLett.86.5341} (for the Hatano-Nelson model, $u$ is a constant). So $P_{t,s}(k_x)$ can still decay rapidly if the non-reciprocity (skin effect) is small compared to the spectral gaps. 

In fact, this criterion extends to more complicated skin effects that can be captured by non-Bloch theory, if we replace the spectral gaps by those corresponding to the PBCs.
For these systems, we work out the band projection using the Green's function. As we only need the decaying behaviour in one direction, we omit the dependence on $k_x$ and reduce the question to one dimension. As before, we use the notation $\mu$ for the shift of crystal momentum into the complex plane.

The projection to a set of bands under open boundary condition is given by integrating the Green's function over frequency \cite{PhysRevB.103.L241408}
\begin{equation}
    P_{t,s}=\oint_{c_\omega} d\omega \oint_{|\beta(\omega)|=e^{\mu(\omega)}} \frac{d\beta}{2\pi i\beta}\beta^{(t-s)}[\omega-h(\beta)]^{-1},\label{eq_pjogr}
\end{equation}
where the expression $h(\beta)$ is to replace the Bloch momentum $e^{ik_y}$ in $H(k_y)$ by $\beta$. The integral for $\omega$ is over a contour $c_\omega$ on the complex plane enclosing the bands, and the integral along $|\beta(\omega)|=e^{\mu(\omega)}$ can be understood as over the generalized Bloch momentum.  This generalized Bloch vector is chosen to be between the $n$-th and $n+1$-th roots of the polynomial $\det[\omega-h(\beta)]$: $|\beta_n(\omega)|<e^{\mu(\omega)}<|\beta_{n+1}(\omega)|$, where $n$ corresponds to the maximal negative powers of $\beta$ in the polynomial $\det[\omega-h(\beta)]$ \cite{kato2013perturbation, PhysRevB.103.L241408}. 

According to our assumption of spectral gaps for both PBC and OBC, we can first fix the contour $c_\omega$ in Eq.~\eqref{eq_pjogr} inside the periodic-boundary spectral gap. For any such energy point $\omega$ outside the periodic-boundary spectrum, one can readily find that the winding number with respect to $\omega$
\begin{align}
    W(\omega)&=\int_0^{2\pi} \frac{dk}{2\pi i} \partial_k \ln \det[\omega-h(k)]\nonumber\\
    &=\oint_{|\beta|=1} \frac{d\beta}{2\pi i } \partial_\beta \ln \det[\omega-h(\beta)]
\end{align}
is trivially zero, namely $W(\omega)=0$. Moreover, the argument principle tells that $W(\omega)=Z-n$, where $Z$ counts how many zeros of the polynomial $\det[\omega-h(\beta)]$ are circled by the conventional BZ with $|\beta|=1$. Consequently, $W(\omega)=0$ implies that the roots of $\det[\omega-h(\beta]=0$ must satisfy $|\beta_n(\omega)|<1<|\beta_{n+1}(\omega)|$. This property ensures that when evaluating the integral in Eq.~\eqref{eq_pjogr}, we can find a positive number $\delta$ such that $|\beta(\omega)|<1-\delta,\forall \omega$ for situation $t-s>0$, and $|\beta(\omega)|>1/(1-\delta), \forall \omega$ for situation $t-s<0$. Eventually, we can bound Eq.~\eqref{eq_pjogr} by
\begin{equation}
    |P_{t,s}|< Z(1-\delta)^{|t-s|},
\end{equation}
where $Z$ is a constant number. This demonstrates the decaying behaviors of $P_{t,s}$ in $|t-s|$. The translational invariance is apparent from Eq.~\eqref{eq_pjogr}.

With the above results, we argue that the imaginary spectral gaps between the band pair under PBCs and OBCs must be opened simultaneously during PT breakings. The total projector $P_{\textrm{sum}}$ to the two bands together is exponentially decaying during the whole real-complex transition, because the spectral gaps to remote bands have nonzero real parts that are assumed to remain nonvanishing during the transition. After the imaginary spectral gap is opened under OBCs, we can decompose the total projector into the projectors to the two individual complex bands $P_{\textrm{sum}}= P_++ P_-$. Here $P_\pm$ projects to the band on the lower/upper half complex plane. If the imaginary spectral gap between the band pair is only opened under OBCs while closed under PBCs, the projector to each complex band will grow in one direction (e.g., $t-s>0$) while decaying in the opposite direction (e.g., $t-s<0$), according to the Green's functions above. Due to PT symmetry, $P_+$ and $P_-$ grow in opposite directions. There is no way that their sum $P_{\textrm{sum}}$ is still exponentially decaying, leading to a contradiction.

\section{Appendix B: Fourier transformation and open boundary Chern-Euler duality}

We give more details about the projection operators and the Chern-Euler duality principle under OBCs. As shown in Appendix A, the band projection operator is decaying quickly and translationally invariant in the bulk up to corrections exponentially small in the systems size $L$. We use the cylinder geometry in the main text as an example. The Fourier transformation of the projection operator is formally defined as 
\begin{align}
    \tilde P(k_x,k_y)=&\sum_{r} P_{r,r_0}(k_x) e^{-ik_y r}\\
    \approx& \sum_{r,s} \frac{1}{L}P_{s+r,s}(k_x) e^{-ik_y r},
\end{align}
where $1 \ll r_0 \ll L$ is a reference position well inside the bulk.
The Fourier transformation is well-defined and continuous in $k_y$ if $\sum_r |P_{r,r_0}|$ converges ($l^1$-integrable) \cite{rudin1987real}. The difference between the first line and the second line comes from two contributions. In the bulk, there are exponentially small corrections. On the boundary, the correction of each $P_{s+r,s}$ is not small, but their contribution is suppressed by the $1/L$ factor. So the two lines agree in the $L\to \infty$ limit. 
In particular, the influence of in-gap boundary modes is eliminated in the $L\to \infty$ limit, allowing $\tilde P$ to be defined unambiguously from the real-space projection. For translational invariant and rapidly decaying $P_{t,s}(k_x)$, the Fourier transformation $\tilde P(\mathbf k)$ is also a projection operator $\tilde P^2(\mathbf k)=\tilde P(\mathbf k)$ \cite{suppm}.

We can now prove that the complex bands split from the real bands carry $|C_\pm|=|\chi|$ under open boundary conditions. We consider generic situations where there can be more than one remote band. The key ingredient we need (Methods and Supplemental Material S4 in \cite{yang2025spontaneous}) is that the sum projection operator $\tilde P_{\textrm{sum}}(\mathbf k)$ to the two bands together varies continuously during spontaneous PT breaking, and decomposes into the summation of two continuous complex conjugate projectors after opening the imaginary spectral gap $\tilde P_{\textrm{sum}}=\tilde P_++\tilde P_-$. The decomposition can be done by taking the contours of the resolvent integrals in Eq.~\eqref{eq:Pdef} symmetric with respect to the real axis after the PT breaking. The continuity of $\tilde P_{\textrm{sum}}(\mathbf k)$ in $(\mathbf k,\lambda)$ can be deduced from the spectral gaps to remote bands and the locality of $P_{\textrm{sum},t,s}$ \cite{suppm}. The continuity of $\tilde P_+,\tilde P_-$ follows similarly. To be more explicit, we can use the image of $\tilde P(\mathbf k)$ to define a set of Bloch states $|\psi(\mathbf k)\rangle$. Such states enables us to study the topology of the eigenstates in the absence of Bloch theory. In particular, the continuity of $\tilde P(\mathbf k)$ leads to the continuity of the vector space spanned by $|\psi(\mathbf k)\rangle$. If we replace the Bloch states in Ref.~\onlinecite{yang2025spontaneous} by the images $|\psi(\mathbf k)\rangle$ of $\tilde P(\mathbf k)$, the Chern-Euler duality rule carries over to the cylinder geometry without any changes. This shows that the resulting two complex bands should carry $C_\pm=\pm|\chi|$ after a spontaneous PT breaking.

The Chern number of $|\psi(\mathbf k)\rangle$ can be extracted via the covariant (biorthogonal) Berry curvature \cite{sun1993high,yang2023homotopy,yang2025spontaneous} dictated by the oblique projection operator $\tilde P(\mathbf k)$ [when the rank of $\tilde P(\mathbf k)$ is one, the Chern number can also be given by the
Hermitian (right-right eigenstate) Berry curvature \cite{PhysRevLett.120.146402,yang2023homotopy}]. By direct calculations, the covariant Berry curvature of $|\psi(\mathbf k)\rangle$ is equal to \cite{PhysRevLett.121.136802}
\begin{equation}
    B(\mathbf k)=\textrm{tr }\tilde P(\mathbf k)[\partial_{k_x}\tilde P(\mathbf k),\partial_{k_y}\tilde P(\mathbf k)].\label{eq_bcp}
\end{equation}

An observation is that the integral of Eq.~\eqref{eq_bcp} can only change when $\tilde P(\mathbf k)$ is discontinuous. The Fourier transformation of a $l^1$-integrable function is always continuous. Therefore, the Chern number can only change when the real-space band projector becomes long-range in the bulk.

 

\renewcommand{\theequation}{S\arabic{equation}}
\setcounter{equation}{0}
\renewcommand{\thefigure}{S\arabic{figure}}
\setcounter{figure}{0}
\renewcommand{\thetable}{S\arabic{table}}
\setcounter{table}{0}
\setcounter{section}{0}
\setcounter{subsection}{0}
\renewcommand{\thesubsection}{S\arabic{subsection}}
\onecolumngrid

\section{Supplemental Materials}

\subsection{Cases when skin effects are absent}

We give the results of numerical calculations of the model of Eqs.\ (\ref{eq:Hlambda})--(\ref{eq:H1}) when the nonreciprocal coupling in $H_1$ is set to zero, $\gamma=0$. In this situation, the system has individual P symmetry and T symmetries, where $P$ is defined such that orbitals 1 and 2 are even under space inversion and orbital 3 is odd, i.e., $H(\mathbf k) = I_P H(-\mathbf k) I_P$, with $I_P = \mbox{diag}\,(1,1,-1)$. 

\begin{figure}
    \centering
    \includegraphics[width=1.0\linewidth]{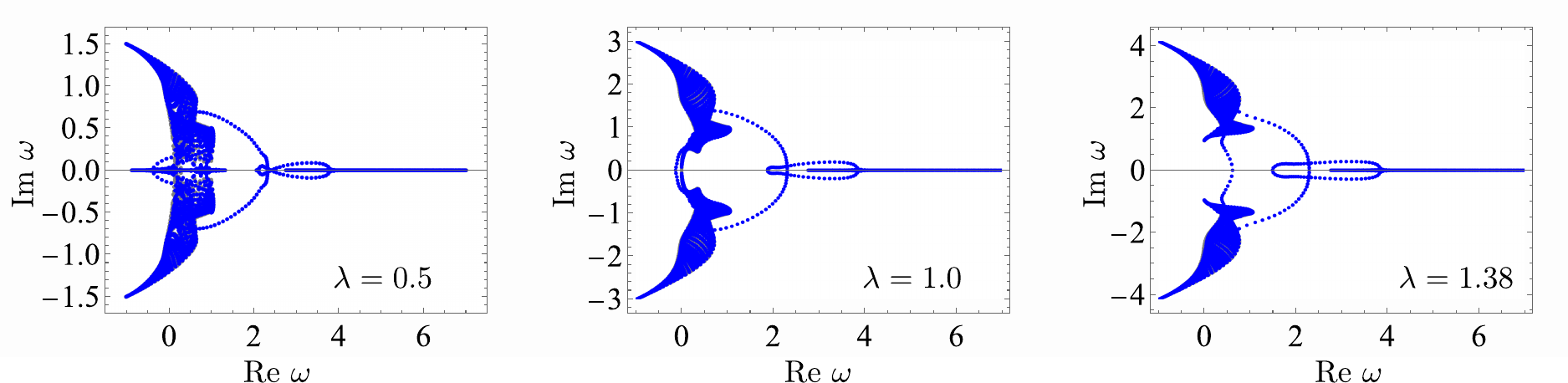}
    \caption{The spectrum of the model of Eqs.\ (\ref{eq:Hlambda})--(\ref{eq:H1}) with $\gamma=0$, which is separately $P$- and $T$-symmetric. The spectrum for open and periodic boundary conditions is shown in blue and grey, respectively. The open boundary bulk spectrum coincides with the periodic boundary bulk spectrum during the entire real-complex transition.}
    \label{fig_IvB}
\end{figure}

The results are shown in Fig.~\ref{fig_IvB}. The situation at $\lambda=0$ is the same as in the main text. The spectra computed for the open boundary condition (cylinder geometry) are shown blue, while the spectra of the same system, computed with periodic boundary conditions (torus geometry) are shown in gray. As one can clearly see from Fig.\ \ref{fig_IvB}), during the spontaneous PT breaking transition induced by the non-Hermitian PT-symmetric couplings, the bulk spectra with open boundary condition and those with the periodic boundary condition coincide, indicating the absence of a skin effect when we have separate P and T symmetries.

The attenuation of skin effects can be explained with the Amoeba formulation. 
The generalized Brillouin zone and the spectrum under the open boundary condition (OBC) are determined by the global minimum of the Ronkin functions defined by
\begin{equation}
R(E,\boldsymbol{\mu})=\int_{T^2}\frac{d^2\boldsymbol{\theta}}{(2\pi)^2}\log \left\vert\det[E-H(\boldsymbol{\theta}-i\boldsymbol{\mu})]\right\vert,\label{eq_rk}
\end{equation}
where $E$ is a complex number and $\boldsymbol{\theta},\boldsymbol{\mu}$ are real two-dimensional variables. The OBC spectrum of the system is obtained in two steps \cite{PhysRevX.14.021011}: First, find out the global minimum of $R(E,\boldsymbol{\mu})$ in $\boldsymbol{\mu}$ at each $E$, denoted as $\boldsymbol{\mu}_{\mathrm{m}}(E)$. Then the density of eigenstates at $E$ is given by the second-order derivative $\rho(E)=\Delta R(E,\boldsymbol{\mu}_{\mathrm{m}}(E))$, where $\Delta=\partial^2/\partial(\mathrm{Re }E)^2+\partial^2/\partial(\mathrm{Im }E)^2$.
Note that in Eq.~\eqref{eq_rk}, the role of the variable $\boldsymbol{\mu}$ is to give $\mathbf k$ an imaginary part, so $\boldsymbol{\mu}_{\mathrm{m}}(E)$ can be intuitively understood as generalizing the conventional Bloch vectors to complex vectors. If the global minimum $\boldsymbol{\mu}_{\mathrm{m}}$ is always reached at $\boldsymbol{\mu}=0$, the generalized Brillouin zone coincides with the conventional Brillouin zone and the OBC spectrum overlaps with the periodic boundary spectrum.

The Ronkin function is a convex function in $\boldsymbol{\mu}$ \cite{PhysRevX.14.021011,ronkinconv}. If a system is invariant under the inversion symmetry, the Bloch Hamiltonian satisfies $H(-\mathbf k)=UH(\mathbf k)U^{-1}$ with $U$ an unitary matrix. For a Hamiltonian $H(\mathbf k)$ that is analytic in $\mathbf k$, this relation also holds when $\mathbf k$ are complex. Thus the inversion symmetry gives $R(E,\boldsymbol{\mu})=R(E,-\boldsymbol{\mu})$, as the determinant is invariant under a unitary transformation and the $\boldsymbol\theta$ integral is unchanged under $\boldsymbol\theta\to -\boldsymbol\theta$. This indicates that for any $\boldsymbol{\mu}\ne 0$, we have $R(E,\boldsymbol{\mu})\ge R(E,0)$, otherwise the line segment joining $\boldsymbol{\mu}$ and $-\boldsymbol{\mu}$ will lie below $R(E,0)$, violating the condition of convex $R(E,\boldsymbol{\mu})$.
So $\boldsymbol{\mu}= 0$ is a global minimum of $R(E,\boldsymbol{\mu})$. Hence, the OBC bulk spectrum overlaps with the Bloch spectrum under periodic boundary condition.


\subsection{Lifting degeneracy of the boundary in-gap modes}

\begin{figure}
    \centering
    \includegraphics[width=1.0\linewidth]{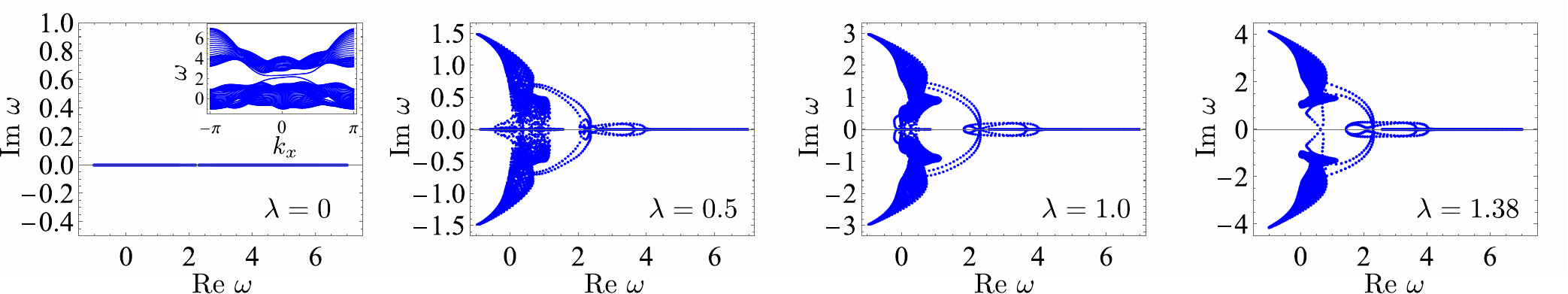}
    \caption{A Hermitian inversion-symmetry breaking coupling (which still respects PT symmetry) can lift the degeneracy between the bottom boundary modes and the top boundary modes. Results are computed at different $\lambda$ for $\gamma=0,\gamma'=0.2$.}
    \label{fig_liftbdegn}
\end{figure}

The degeneracy between the top and bottom boundary in-gap modes can be lifted by including an Hermitian inversion-symmetry breaking coupling $\sum_{\mathbf r}\gamma'(|\mathbf r,1\rangle\langle\mathbf r,3|+|\mathbf r,3\rangle\langle\mathbf r,1|)$ ($\gamma\in \mathbb R$) between the orbitals 1,3. This term still preserves the PT symmetry as its coefficient is real. We compute the results for $\gamma'=0.2,\gamma=0$. The results are shown in Fig.~\ref{fig_liftbdegn}.

The Hermitian inversion-breaking term makes the spectrum no longer symmetric with respect to $k_x=0$, as in the inset of the first figure in Fig.~\ref{fig_liftbdegn}. After the PT breaking, the top  boundary modes and the bottom boundary modes split. Each boundary carries two in-gap modes and those at opposite boundaries are related by a complex conjugation transformation (PT reversal).

\subsection{Calculation for the flow of state}

We expand the expression for $\delta n_{in}$. With the commutator, we have
\begin{align}
    -\delta n_{in}=\sum_{t,s}\textrm{tr}\left([P,\dot P]_{s,t}[P,Q]_{t,s}\right)=\sum_{t,s}\textrm{tr }[P,\dot P]_{s,t}P_{t,s}[\theta(s-y_1)\theta(y_2-s)-\theta(t-y_1)\theta(y_2-t)].\label{eq_flexp}
\end{align}
The differences between the products of the Heaviside functions are nonzero when: (i) $y_1<s<y_2$, $t<y_1$ or $t>y_2$. (ii) $y_1<t<y_2$, $s<y_1$ or $s>y_2$. Assuming that the decaying length scale of $|P_{s,t}P_{t,s}(t-s)|$ is $R$, we see that Eq.~\eqref{eq_flexp} also vanishes when $|s-y_j|>R$ or $|t-y_j|>R$ all $j$. So we can restrict the summation in Eq.~\eqref{eq_flexp} to $s,t\in [y_1-R,y_1+R]$ and $s,t\in [y_2-R.y_2+R]$. Here we assume $|y_1-y_2|\gg R$. For $s,t\in [y_1-R,y_1+R]$, Eq.~\eqref{eq_flexp} becomes:
\begin{equation}
    \sum_{t,s\in [y_1-R,y_1+R]}\textrm{tr }[P,\dot P]_{s,t}P_{t,s}[\theta(s-y_1)-\theta(t-y_1)].
\end{equation}
This is the expression for $J_{y_1}$. The expression for $J_{y_2}$ is obtained in a similar way by replacing $\theta(x)=1-\theta(-x-\epsilon)$, where $\epsilon$ is an infinitesimal positive number. 

\subsection{Fourier transformation of a translationally invariant local projection is also a projection}

Assume that $P_{t,s}$ is a translationally invariant projection and decays quickly as a function of $|t-s|$. We use $\tilde P(k)$ for its Fourier transformation:
\begin{equation}
    \tilde P(k)=\sum_r P_{r,r_0}e^{-ikr},
\end{equation}
where as in the main text, we omit the orbital indices and $\tilde P(k)$ is a smaller matrix compared to $P_{t,s}$. The square of $\tilde P(k)$ is
\begin{equation}
    \tilde P^2(k)=\sum_{r,r'} P_{r,r'}e^{-ikr}P_{r',r_0}e^{-ikr'}=\sum_{r,r'} P_{r+r',r'}P_{r',r_0}e^{-ik(r+r')}=\sum_{\bar r=r+r'} P_{\bar r,r_0}e^{-ik\bar r}=\tilde P(k).
\end{equation}
In the second step, we use the translational invariance of $P_{t,s}$. This replacement is valid except for corrections from the boundaries, which is small in the limit $L\to\infty$. In the last step, we denote $r+r'$ as $\bar r$ and use the projection condition $\sum_r' P_{\bar r,r'}P_{r',r_0}=P_{\bar r,r_0}$. These equations are valid when $P_{t,s}$ decays quickly in $|t-s|$. This verifies that $\tilde P(k)$ is also a projection.

\subsection{Continuity of the Fourier transformation of the real-space projection operator}

In this part, we consider a real-space local Hamiltonian $H_{t,s}(\lambda)$ that is continuous on a set of parameters $\lambda$. We show that the Fourier transformation $\tilde P(k,\lambda)$ of the real-space band projection $ P_{t,s}(\lambda)$ is continuous in $(k,\lambda)$. By replacing $k\to k_y$, $\lambda\to (k_x,\lambda)$, this will prove the continuity of $\tilde P(\mathbf k)$ in the main text during the real-complex transition. 

We assume that the $l^1$-sum of $P_{t,s}(\lambda)$ has a upper bound $\sup_{\lambda}\sum_r|P_{r,r_0}(\lambda)|=S<\infty$ and the $l^1$-convergence is reached within a radius $R(\delta)$ such that $\sup_\lambda\sum_{|r-r_0|>R(\delta)}|P_{r,r_0}(\lambda)|<\delta$ for arbitrary $\delta$. These assumptions follow from the spectral gaps between the bands that we study and the other bands. We estimate the variation between the Fourier transformations
\begin{align}
    |\tilde P(k',\lambda')-\tilde P(k,\lambda)|&=|\tilde P(k',\lambda')-\tilde P(k',\lambda)+\tilde P(k',\lambda)-\tilde P(k,\lambda)|\nonumber\\
    &\le\left|\sum_r[P_{r,r_0}(\lambda')-P_{r,r_0}(\lambda)]e^{ik'r}\right\vert+\left\vert\sum_r P_{r,r_0}(\lambda)(e^{-ikr}-e^{-ik'r})\right\vert\nonumber\\
    &\le\sum_r|P_{r,r_0}(\lambda')-P_{r,r_0}(\lambda)|+\sum_r \left\vert P_{r,r_0}(\lambda)(e^{-ikr}-e^{-ik'r})\right\vert.\label{eq_pcon}
\end{align}
Since $\sum_r|P_{r,r_0}(\lambda')-P_{r,r_0}(\lambda)|\le \Vert P(\lambda')-P(\lambda)\Vert$ [$P(\lambda)$ as a matrix in the real-space coordinates], we bound the first term in Eq.~\eqref{eq_pcon} by
\begin{align}
    \Vert P(\lambda')-P(\lambda)\Vert=&\left\Vert \frac{1}{2\pi i} \oint_{\mathcal C_\alpha} dz[z-H(\lambda')]^{-1}[H(\lambda')-H(\lambda)][z-H(\lambda)]^{-1}\right\Vert\nonumber\\
    \le &\frac{1}{2\pi} \oint_{\mathcal C_\alpha} |dz|\left\Vert[z-H(\lambda')]^{-1}\right\Vert \left\Vert[z-H(\lambda)]^{-1}\right\Vert \left\Vert H(\lambda')-H(\lambda)\right\Vert.
\end{align}
Due to the spectral gaps, $\left\Vert[z-H(\lambda')]^{-1}\right\Vert$ and $\left\Vert[z-H(\lambda)]^{-1}\right\Vert$ are upper bounded. The above equation goes to zero as $(k',\lambda')\to(k,\lambda)$ because of the continuity and locality of $H(\lambda)$. For the second term in Eq.~\eqref{eq_pcon}, we divide it into \begin{align}
    \sum_r \left\vert P_{r,r_0}(\lambda)(e^{-ikr}-e^{-ik'r})\right\vert=&\sum_{|r-r_0|\le R(\delta)} \left\vert P_{r,r_0}(\lambda)(e^{-ikr}-e^{-ik'r})\right\vert+\sum_{|r-r_0|>R(\delta)}\left\vert P_{r,r_0}(\lambda)(e^{-ikr}-e^{-ik'r})\right\vert\nonumber\\
    \le & SR(\delta)|k-k'|+2\delta.
\end{align}
By choosing $\delta=\epsilon/3$ and $|k-k'|<\epsilon/SR(\epsilon/3)$, the above equation can be smaller than arbitrary $\epsilon$. Thus, the second term of Eq.~\eqref{eq_pcon} also goes to zero as $(k',\lambda')\to (k,\lambda)$. This concludes the proof for the continuity of $\tilde P(k,\lambda)$.

\end{document}